\documentclass[pra,twocolumn,amsmath]{revtex4-1}

\usepackage{graphicx}
\usepackage{color}

\begin{document}

\title{Time delays in ultracold atomic and molecular collisions}

\author{Matthew D. Frye}

\affiliation{Joint Quantum Centre (JQC) Durham/Newcastle, Department of
Chemistry, Durham University, South Road, Durham DH1 3LE, United Kingdom}

\author{Jeremy M. Hutson}
\affiliation{Joint Quantum Centre (JQC) Durham/Newcastle, Department of
Chemistry, Durham University, South Road, Durham DH1 3LE, United Kingdom}

\date{\today}

\begin{abstract}
We study the behavior of the Eisenbud-Wigner collisional time delay around
Feshbach resonances in cold and ultracold atomic and molecular collisions. We
carry out coupled-channels scattering calculations on ultracold Rb and Cs
collisions. In the low-energy limit, the time delay is proportional to the
scattering length, so exhibits a pole as a function of applied field. At high
energy, it exhibits a Lorentzian peak as a function of either energy or field.
For narrow resonances, the crossover between these two regimes occurs at an
energy proportional to the square of the resonance strength parameter
$s_\textrm{res}$. For wider resonances, the behavior is more complicated and we
present an analysis in terms of multichannel quantum defect theory.
\end{abstract}

\maketitle

\section{Introduction}

Scattering resonances are important in many fields, from nuclear
physics to physical chemistry. A resonance occurs when a collision occurs at an
energy close to that of a quasibound state of the collision complex, so that
scattering flux is temporarily trapped at short range. Resonant scattering is
different in character from non-resonant scattering, and often produces
different products and characteristic angular distributions; it typically
produces strong features in the dependence of collision properties on energy
and external fields. There are particularly important applications in ultracold
atomic physics, where magnetically tunable Feshbach resonances are used to
control the behavior of ultracold atoms.

The language of resonant scattering is often used to understand physical
phenomena. A quasibound state has a width that depends on its coupling to
energetically open channels, and the width is interpreted as inversely
proportional to the lifetime of the state. Conversely, resonant collisions
experience a resonant time delay, which is also usually supposed to be
inversely proportional to the resonance width. However, as will be seen below,
this simple viewpoint breaks down in the low-energy regime close to threshold.

A topical application of resonant time delays, which motivated the current
study, is in collisions of ultracold molecules. If the interaction potential
for a colliding pair has a deep well, the collision complex can have a high
density of states even at low collision energies \cite{Mayle:2012, Mayle:2013,
Christianen:density:2019}. The resulting dense pattern of scattering resonances
may produce long-lived ``sticky'' collisions in the ultracold regime. The
collision complexes may be destroyed either by collision with a third body
\cite{Mayle:2013} or by laser-driven processes \cite{Christianen:laser:2019}.
If the lifetime of the complex is long compared to the destruction mechanism,
the overall process displays second-order kinetics. Multiple experiments have
reported short trap lifetimes for molecules that have no 2-body collisional
loss mechanism \cite{Takekoshi:RbCs:2014, Molony:RbCs:2014, Park:NaK:2015,
Guo:NaRb:2016, Ye:2018}, which may be a sign of such effects. Gregory \emph{et
al.}\ \cite{Gregory:RbCs-collisions:2019} have demonstrated that the kinetics
of the loss process are indeed second order for ultracold RbCs.
It is thus important to understand collisional time delays
in the ultracold regime.

The theory of collisional time delays in quantum scattering was established by
Eisenbud, \cite{Eisenbud:dissertation:1948}, Wigner \cite{Wigner:time:1955} and
Smith \cite{Smith:1960}. It has been used to analyse resonant contributions to
recombination at non-ultracold temperatures \cite{Kendrick:1995,
Kendrick:1996}. In a few cases time delays have been calculated for ultracold
scattering \cite{Field:2003, Guillon:2009, Bovino:2011, Simoni:2009,
Croft:2017, Mehta:2018}. However, there has been little work on understanding
the basic properties and behaviors of the time delay in ultracold collisions.
The purpose of the present paper is to explore how time delays behave close to
threshold. We will show that, in this regime, the time delay does not show a
simple peak around a resonance. The behavior is particularly striking when
viewed as a function of external field rather than energy: as a function of
field, the time delay may be either positive or negative, and in the low-energy
limit averages to zero across a resonance. We will illustrate the behavior with
calculations on resonances in ultracold atomic collisions, and discuss the
transition from the threshold regime to higher energy.

\section{Eisenbud-Wigner-Smith Time Delay}

Eisenbud \cite{Eisenbud:dissertation:1948} and Wigner \cite{Wigner:time:1955}
used a wavepacket analysis to define a time delay for
single-channel scattering,
\begin{equation}
Q(E)=2\hbar\frac{d\delta}{dE},
\label{eq:Q}
\end{equation}
where $\delta$ is the scattering phase shift and $E$ is the energy. Smith
\cite{Smith:1960} considered the problem in a time-independent formalism and
defined a time-delay matrix suitable for multichannel scattering,
\begin{equation}
\boldsymbol{Q}(E)=i\hbar\boldsymbol{S}\frac{d\boldsymbol{S}^\dagger}{dE},
\label{eq:Q_matrix}
\end{equation}
in terms of the scattering matrix $\boldsymbol{S}$. If there is only a single
open channel, $\boldsymbol{S}=e^{2{\rm i}\delta}$ and Eq.\ \eqref{eq:Q_matrix}
reduces to Eq.\ \eqref{eq:Q}. The present work will focus on the case of a
single open channel, but will consider Feshbach resonances
due to the effects of additional closed channels.

\subsection{Far above threshold}

We first consider an isolated narrow resonance far above threshold. The elastic
scattering phase shift follows a Breit-Wigner form as a function of energy at
constant field,
\begin{equation}
\delta(E)=\delta_{\rm bg}(E)+\arctan \left[\frac{\frac{1}{2}
\Gamma_\textrm{E}(E)}{E_\textrm{res}-E}\right],
\label{eq:delta_res}
\end{equation}
where $\delta_{\rm bg}(E)$ is a background phase shift that is a slow function
of energy, $E_\textrm{res}$ is the resonance energy, and $\Gamma_\textrm{E}(E)$
is the resonance width in energy. The phase shift increases by $\pi$ above its
background value across the width of a resonance. Far above threshold, the
dependence of $\Gamma_\textrm{E}$ on $E$ can usually be neglected, and the time
delay is \cite{Smith:1960}
\begin{equation}
Q(E)=Q_{\rm bg}(E) + \frac{\hbar\Gamma_\textrm{E}}{(E_\textrm{res}-E)^2+\Gamma_\textrm{E}^2/4}.
\label{eq:Q_res}
\end{equation}
This shows a simple Lorentzian peak as a function of energy.
Neglecting the background term, the integral across this peak is $2\pi\hbar$,
independent of $\Gamma_\textrm{E}$. An important consequence of this is that,
if $\Gamma_\textrm{E}(E) \ll k_{\rm B}T$, the resonant contribution to a
thermally averaged time delay is independent of the width of the resonance.
Thus, under some circumstances, the contribution of a large number of narrow
resonances can be understood from the density of states without any more
detailed understanding of the interactions and dynamics. An approximation of
this form was used by Bowman \cite{Bowman:1986} to obtain an alternative
derivation of RRKM theory.

For cold collisions it is common to consider the resonance as a function of
external field (here taken to be magnetic field $B$), at a constant collision
energy $E$ with respect to a (potentially field-dependent) threshold energy
$E_\textrm{thresh}$. The phase shift is given by
\begin{equation}
\delta(E,B)=\delta_\textrm{bg}(E)+\arctan
\left[\frac{\frac{1}{2}\Gamma_\textrm{B}(E)}{B-B_\textrm{res}^\textrm{BW}(E)}\right].
\label{eq:delta_res_B}
\end{equation}
Here $B_\textrm{res}^\textrm{BW}(E)$ is the position of resonance in field. Far
from threshold $B_\textrm{res}^\textrm{BW}(E)$ varies with energy according to
the magnetic moment of the resonant state relative to the threshold,
$\mu_\textrm{rel} = d(E_\textrm{bound}-E_\textrm{thresh})/dB$. The resonance
width in field is $\Gamma_\textrm{B}(E)=\Gamma_\textrm{E}(E)/\mu_\textrm{rel}$.
The time delay can then be written
\begin{equation}
Q(E,B)=Q_{\rm bg}(E) + \frac{\hbar\Gamma_\textrm{B}(E)/\mu_\textrm{rel}}
{[B_\textrm{res}^\textrm{BW}(E)-B]^2+\Gamma_\textrm{B}(E)^2/4}.
\label{eq:Q_res_B_high}
\end{equation}
Far above threshold, the time delay thus shows a
Lorentzian peak as a function of external field as well as energy.

\subsection{Ultracold scattering}

In the ultracold regime, scattering is modified by threshold effects
\cite{WIGNER:1948}. These are conveniently expressed in terms of the wavenumber
$k$, where $E=\hbar^2 k^2/2\mu$ and $\mu$ is the collisional reduced mass. A
key quantity is the $k$-dependent scattering length,
\begin{equation}
a(k,B)=\frac{-\tan\delta(k,B)}{k}.
\end{equation}
For some purposes it is sufficient to consider only the zero-energy scattering
length, $a(B)=\lim_{k\to 0}a(k,B)$; in the low-energy limit, $\delta=-ka(B)$,
and \cite{Field:2003}
\begin{equation}
Q=\frac{-2a(B)\mu}{\hbar k}=-2\frac{a(B)}{v}
\label{eq:Q_a}
\end{equation}
where $v=k\hbar/\mu$ is the collision velocity. This is exactly the classical
time delay associated with a hard-sphere collision with radius $a(B)$
\cite{Simoni:2009}, in accordance with the usual interpretation of the
scattering length.

Around a low-energy Feshbach resonance, the scattering length shows a pole as a
function of field \cite{Moerdijk:1995},
\begin{equation}
a(k,B)=a_\textrm{bg}(k)\left[1-\frac{\Delta(E)}{B-B_\textrm{res}^\textrm{pole}(E)}\right],
\label{eq:a_B}
\end{equation}
where $\Delta(E)$ characterizes the width of the pole. The pole position
$B_\textrm{res}^\textrm{pole}(E)$ coincides with
$B_\textrm{res}^\textrm{BW}(E)$ at zero energy, but they generally differ away
from threshold, as discussed in section \ref{sec:qdt}. As the
scattering length passes through both large positive and large negative values
near the pole, Eq.\ \eqref{eq:Q_a} implies that there are both positive and
negative time delays around a resonance in the low-energy limit
\cite{vanDijk:1992,Field:2003}.
This behavior is very different from that seen away from threshold, Eq.\
\eqref{eq:Q_res}, where the resonant contribution to the time
delay is strictly positive.

\subsection{Intermediate regime}

In order to reconcile the different behavior of $Q$ close to threshold and far
from it, it is necessary to consider the energy dependence of the resonance
parameters. The phase shift can still be written in the form
\eqref{eq:delta_res_B}, but the derivatives of the parameters with respect to
energy are important. The full expression for the time delay is
\begin{align}
Q(E,B)&=Q_\textrm{bg}(E) \nonumber \\
&+\frac{d\Gamma_\textrm{B}}{dE}\frac{\hbar[B-B_\textrm{res}^\textrm{BW}(E)]}{[B-B_\textrm{res}^\textrm{BW}(E)]^2+\Gamma_\textrm{B}(E)^2/4} \nonumber \\
&+\frac{dB_\textrm{res}^\textrm{BW}}{dE}\frac{\hbar\Gamma_\textrm{B}(E)}{[B-B_\textrm{res}^\textrm{BW}(E)]^2+\Gamma_\textrm{B}(E)^2/4}. \label{eq:Q_full}
\end{align}
In the high-energy limit, $d\Gamma_\textrm{B}/dE=0$ and Eq.\ \eqref{eq:Q_full}
reduces to Eq.\ \eqref{eq:Q_res_B_high}. In the low-energy limit, the energy
dependence of the width can be written
$\Gamma_\textrm{B}(E)=2ka_\textrm{bg}\Delta$ \cite{Mies:Feshbach:2000};
$d\Gamma_\textrm{B}/dE$ diverges as $k\rightarrow 0$, so the second term in
Eq.\ \eqref{eq:Q_full} dominates.

The crossover between these limiting behaviors occurs around a crossover energy
$E_\textrm{X}$ where the two derivatives in Eq.\ \eqref{eq:Q_full} are equal.
As described below, the threshold has little effect on
$B_\textrm{res}^\textrm{BW}(E)$ when the resonance is narrow or $a_\textrm{bg}$
is close to $\bar{a}$, where $\bar{a} = 0.4779888\dots \times (2\mu
C_6/\hbar^2)^{1/4}$ is the mean scattering length of Gribakin and Flambaum
\cite{Gribakin:1993} for an interaction potential $-C_6R^{-6}$.
$dB_\textrm{res}^\textrm{BW}/dE$ is thus approximately $1/\mu_\textrm{rel}$,
and the low-energy expression for $\Gamma_\textrm{B}(E)$ gives
\begin{equation}
E_\textrm{X} \approx \frac{2\mu}{\hbar^2}a_\textrm{bg}^2\Delta^2 \mu_\textrm{rel}^2 = s_\textrm{res}^2\bar{E}. \label{eq:crossover}
\end{equation}
Here, the dimensionless resonance strength parameter \cite{Chin:RMP:2010} is
$s_\textrm{res}=a_\textrm{bg}\Delta \mu_\textrm{rel}/(\bar{a}\bar{E})$, where
$\bar{E}=\hbar^2/2\mu\bar{a}^2$. Substituting back into Eq.\ \eqref{eq:Q_full},
at $E=E_\textrm{X}$ we expect the peak in $Q$ to be about 20 times larger than
the trough for a narrow resonance.

\section{Numerical examples}

\begin{table*}[htbp]
  \centering
\caption{\label{table:res_params} Parameters for the example resonances,
obtained from coupled-channels calculations.}
\begin{ruledtabular}
    \begin{tabular}{ccccccccccc}
   Res.	& Isotope		& $\bar{a}/a_0$	& $\bar{E}/k_\textrm{B}$ ($\mu$K)	& $B_\textrm{res}^\textrm{pole}(0)$ (G)	& $a_\textrm{bg}/a_0$	& $\Delta$ (G)	& $\mu_\textrm{rel}/\mu_\textrm{B}$		& $s_\textrm{res}$	& $E_\textrm{X}/k_\textrm{B}$ ($\mu$K)	& $\bar{\Gamma}_\textrm{B}$ (mG)	\\
\hline
1		& $^{87}$Rb	& 79.0		& 319						& 1007.86				& 100					& 0.20		& 2.8							& 0.15			& 7.2								& 476	\\
2		& $^{87}$Rb	& 79.0		& 319						& 686.60				& 100					& 0.0072		& 1.3							& 0.0025			& 0.0020							& 17		\\
3		& $^{133}$Cs	& 96.6		& 140						& 47.79				& 1008					& 0.15		& 1.2							& 0.88			& 110							& 35		\\
4		& $^{85}$Rb	& 78.5		& 331						& 851.3				& $-390$					& $-1.2$		& 2.1							& 2.6				& 2200							& 324	\\
\end{tabular}
\end{ruledtabular}
\end{table*}

We illustrate the behavior with calculations on resonances in collisions of Rb
and Cs atoms. We carry out coupled-channels calculations to evaluate
energy-dependent phase shifts in magnetic fields using the \textsc{molscat}
\cite{molscat:2019, mbf-github:2019} package. The methods used are similar to
those described in Ref.\ \onlinecite{Blackley:85Rb:2013}. We use the
interaction potentials of Strauss \emph{et al.}\ \cite{Strauss:2010}
\footnote{The calculations of Blackley et al. \cite{Blackley:85Rb:2013}
included a retardation correction in the long-range part of the interaction
potential, so gave slightly different pole positions.} for Rb and Berninger
\emph{et al.}\ \cite{Berninger:Cs2:2013} for Cs. The energy derivatives
required for the time delay are calculated by finite difference from two
calculations at energies that differ by 0.1\%.

We take four resonances in ultracold Rb and Cs scattering as examples: (1) The
resonance near 1007 G for $^{87}$Rb \cite{Marte:2002, Volz:2003,
Durr:diss:2004}; (2) the resonance near 687 G for $^{87}$Rb \cite{Marte:2002,
Durr:diss:2004}; (3) the resonance near 47 G for $^{133}$Cs
\cite{Chin:cs2-fesh:2004, Lange:2009}; and (4) the resonance near 850 G for
$^{85}$Rb \cite{Blackley:85Rb:2013}. All these resonances are for atoms in
their lowest hyperfine and Zeeman state, so we do not need to consider effects
of inelastic scattering. Scattering involving non-s-wave open channels is
negligible, so we consider only one open channel and use Eq.\ \eqref{eq:Q}. For
each resonance we find $B^\textrm{pole}_\textrm{res}(0)$, $a_\textrm{bg}$, and
$\Delta$ from coupled-channels calculations by converging on and characterizing
the pole in scattering length using the methods of Frye and Hutson
\cite{Frye:resonance:2017}. To obtain $\mu_\textrm{rel}$, we carry out
coupled-channels calculations of the energy of the bound state in a near-linear
region below threshold, using the \textsc{bound} package
\cite{bound+field:2019,mbf-github:2019}. Table \ref{table:res_params} lists
these parameters, together with other relevant quantities including
$s_\textrm{res}$ and $E_\textrm{X}$, for each of the resonances. Resonance 1 is
the widest known for ground-state $^{87}$Rb, but is of only moderate width
compared to resonances in other similar systems; its background scattering
length is close to $\bar{a}_\textrm{87Rb}$. Resonance 2 is significantly
narrower and has essentially the same background scattering length. Resonance 3
has a similar width $\Delta$ to resonance 1, but a much larger background
scattering length, $a_\textrm{bg} \approx 10\bar{a}_\textrm{133Cs}$, so that
$s_\textrm{res}$ is significantly larger. Resonance 4 is a broad resonance with
a large negative scattering length, $a_\textrm{bg} \approx -5
\bar{a}_\textrm{85Rb}$.

\begin{figure*}[htb]
\includegraphics[width=0.95\linewidth]{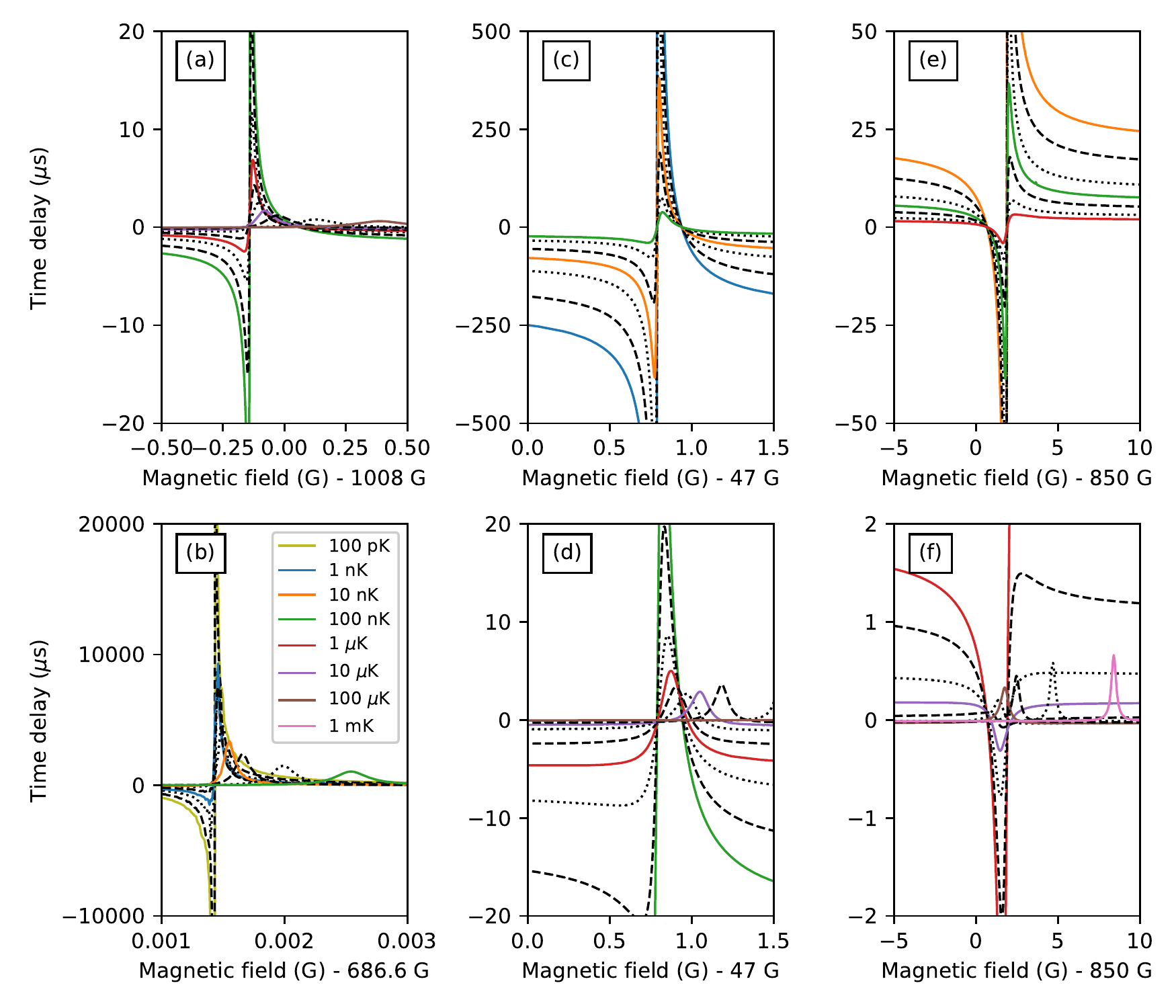}
\caption{\label{fig:plot_grid} Time delay $Q$ as a function of magnetic field
$B$ at various energies for the four example resonances. Dashed and dotted
lines show energies 2 and 5 times those of the corresponding solid lines. (a)
resonance 1; (b) resonance 2; (c) and (d) resonance 3; (e) and (f) resonance
4.}
\end{figure*}

Figure \ref{fig:plot_grid} shows the time delay for the example resonances as a
function of magnetic field for a variety of energies from 100 pK up to 1 mK.
Figure \ref{fig:plot_grid}(a) shows the time delay for resonance 1, for which
$E_\textrm{X}/k_\textrm{B} = 7.2$ $\mu$K. At the lowest energy
shown, 100 nK, it has a large symmetrical pole-like oscillation and deviates
from this only in a small region near the center. The difference in behavior
between the lowest few energies in the wings of the resonance is mostly due to
the dependence on $k$ in Eq.\ \eqref{eq:Q_a}. The pole-like behavior is
suppressed in a region near the center due to the denominator in the second
term of Eq.\ \eqref{eq:Q_full}; the width of this region is proportional to
$\Gamma_\textrm{B}(E)$, so it broadens as the energy increases, greatly
reducing the magnitude of the peak and trough. The importance of the second
term of Eq.\ \eqref{eq:Q_full} decreases with increasing energy; by 1 $\mu$K
the oscillation is significantly asymmetric, and by 10 $\mu$K the trough has
disappeared entirely, leaving a single peak that starts to shift away from the
zero-energy resonance position. This agrees well with the crossover energy
$E_\textrm{X}/k_\textrm{B} = 7.2$ $\mu$K predicted by Eq.\
\eqref{eq:crossover}. The peak then continues to move off to high field and
broaden towards its high-energy form.

Figure \ref{fig:plot_grid}(b) shows the behavior around resonance 2, for which
$E_\textrm{X}/k_\textrm{B}=2$ nK. It shows similar features to resonance 1, but
they occur at much lower energy. The oscillation is already highly asymmetric
at 1 nK. By 10 nK the trough has disappeared and $Q$ reaches its
high-energy form well below 100 nK.

Figures \ref{fig:plot_grid}(c) and (d) show the behavior around resonance 3. As
for resonances 1 and 2, the oscillation is pole-like and symmetric at the
lowest energies, but it develops significant asymmetry by 100 nK, which is far
below the crossover energy predicted by Eq.\ \eqref{eq:crossover},
$E_\textrm{X}/k_\textrm{B} \approx 110\ \mu$K. The shape of $Q$ has become a
single peak by 1 $\mu$K. Above 5 $\mu$K the peak shifts to higher field and
gets narrower and higher.

Figures \ref{fig:plot_grid}(e) and (f) show the behavior around resonance 4. In
this case, the transition from low-energy to high-energy behavior is more
complicated. The oscillation becomes asymmetric around 1 $\mu$K and it is the
trough that is initially more pronounced than the peak. As for
resonance 3, this happens well below the crossover energy predicted by Eq.\
\eqref{eq:crossover}, $E_\textrm{X}/k_\textrm{B} \approx 2.2$ mK. By 10 $\mu$K,
$Q$ has just a single trough with no visible peak, but by 100 $\mu$K this has
inverted to a single peak with no trough. Above 100 $\mu$K, the peak shifts
away to higher field and, as for resonance 3, gets narrower and higher.

The behavior of the time delay for resonances 1 and 2 follows the simple theory
described in section 2. However, the approximate forms of the resonance
parameters used in deriving Eq.\ \eqref{eq:crossover} are not valid for a broad
resonance with $a_\textrm{bg}$ far from $\bar{a}$. Understanding the more
complicated behavior for resonances 3 and 4 show needs a more complete
description of the threshold effects. This is given in the following section.

\section{Interpretation using multichannel quantum defect theory}
\label{sec:qdt}

Multichannel quantum-defect theory (MQDT) provides a unified framework for
describing scattering both close to and far from threshold. We use a 2-channel
MQDT model of the resonance in the formalism of \citeauthor{Mies:MQDT:2000}
\cite{Mies:MQDT:2000}, as described in detail by \citeauthor{Jachymski:2013}
\cite{Jachymski:2013}. This model accurately reproduce the coupled-channels
results. In the model, the width and position of a resonance can be written as
\begin{align}
\Gamma_\textrm{B}(E)&=\bar{\Gamma}_\textrm{B}C^{-2}(E)
\label{eq:QDT_Gamma} \\
B_\textrm{res}^\textrm{BW}(E)
&=B_0+\frac{E}{\mu_\textrm{rel}}-\frac{1}{2}\bar{\Gamma}_\textrm{B}\tan\lambda(E). \label{eq:QDT_Bres}
\end{align}
Here, $\bar{\Gamma}_\textrm{B}$ is the short-range width in field, neglecting
threshold effects, which is independent of both $E$ and $B$, and $B_0$ is the
field at which the bare (uncoupled) bound state crosses threshold. They are
related to $\Delta$ and $B_\textrm{res}^\textrm{pole}(0)$ by
\cite{Julienne:2006}
\begin{align}
\frac{1}{2}\bar{\Gamma}_\textrm{B}&= \frac{r_\textrm{bg}}{1+(1-r_\textrm{bg})^2} \Delta \\
B_0 &= B_\textrm{res}^\textrm{pole}(0) - \frac{r_\textrm{bg} (1-r_\textrm{bg})}{1+(1-r_\textrm{bg})^2}\Delta,
\label{eq:QDT_B0}
\end{align}
where $r_\textrm{bg}=a_\textrm{bg}/\bar{a}$.

The QDT functions $C(E)$ and $\tan\lambda(E)$ were defined by Mies
\cite{Mies:1984a}; $C(E)$ describes the amplitude of the wavefunction at short
range compared to long range, while $\tan\lambda(E)$ describes the modification
in phase due to threshold effects. For a particular long-range potential form,
they are functions of $E/\bar{E}$ that depend parametrically on
$a_\textrm{bg}/\bar{a}$. They can be calculated numerically for arbitrary
potentials \cite{Yoo:1986, Croft:MQDT:2011}. However, in this work we
approximate the potential by its leading dispersion interaction $-C_6R^{-6}$
and use Gao's analytic solutions \cite{Gao:C6:1998} to calculate the QDT
functions. In the QDT model, the phase shift and the time delay are still given
by Eqs.\ \eqref{eq:delta_res_B} and \eqref{eq:Q_full}.

\begin{figure}[htbp]
\includegraphics[width=0.95\linewidth]{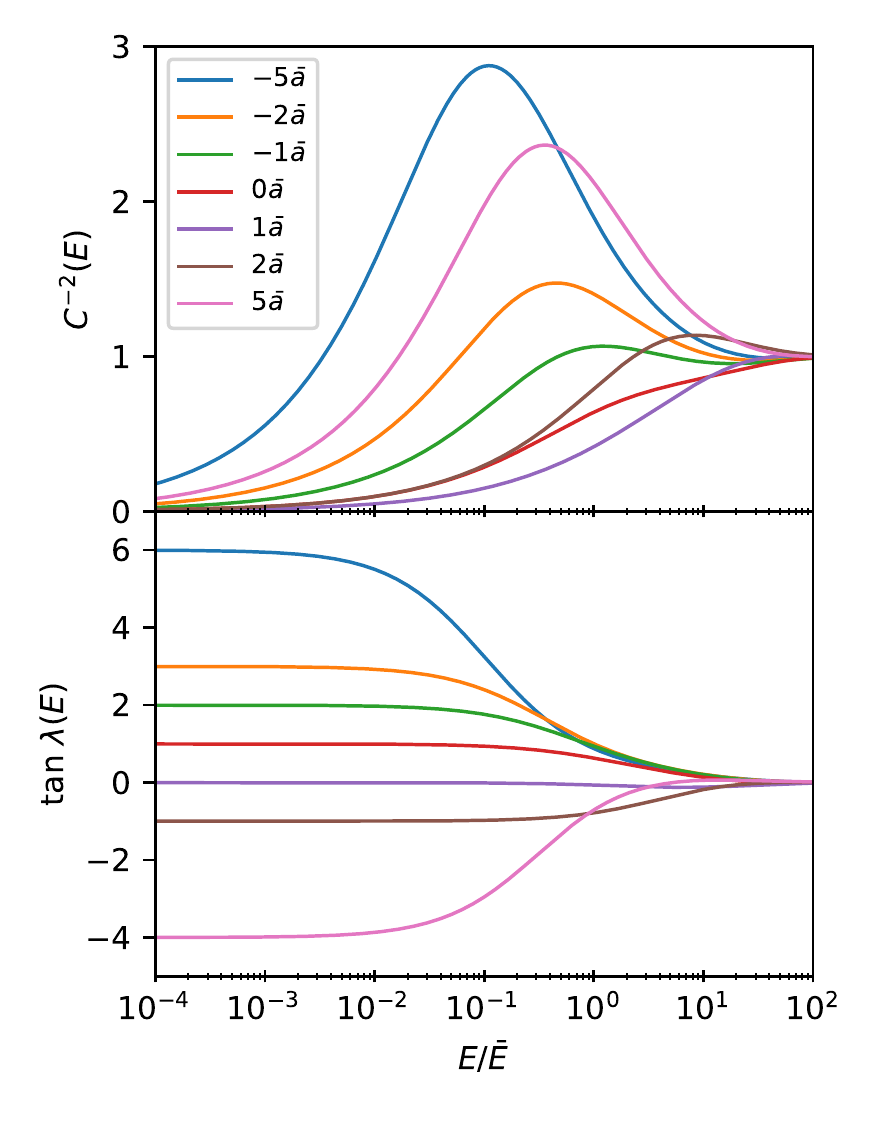}
\caption{\label{fig:qdt_params} QDT functions $C^{-2}(E)$ and
$\tan\lambda(E)$ for various background scattering lengths $a_\textrm{bg}$.}
\end{figure}

Examples of $C^{-2}(E)$ and $\tan\lambda(E)$ are shown in Fig.\
\ref{fig:qdt_params} for a variety of values of $a_\textrm{bg}$. The functions
$C^{-2}(E)$ approach 1 at high energy, but at low energy the leading term is
$k\bar{a}[1+(1-r_\textrm{bg})^2]$, so is the same for $a_\textrm{bg}=0$ and
$2\bar{a}$. For larger values of $|r_\textrm{bg}|$, $C^{-2}(E)$ rises more
rapidly and has a prominent peak; for $|r_\textrm{bg}| \le 1$, this peak is
small or absent. The functions $\tan\lambda(E)$ are $1-r_\textrm{bg}$ at low
energy and approach zero at high energy; they start to decrease at
substantially lower energies for larger values of $|r_\textrm{bg}|$. For
$a_\textrm{bg}=\bar{a}$, $\tan\lambda(E)$ remains small at all energies.

\begin{figure}[htbp]
\includegraphics[width=0.95\linewidth]{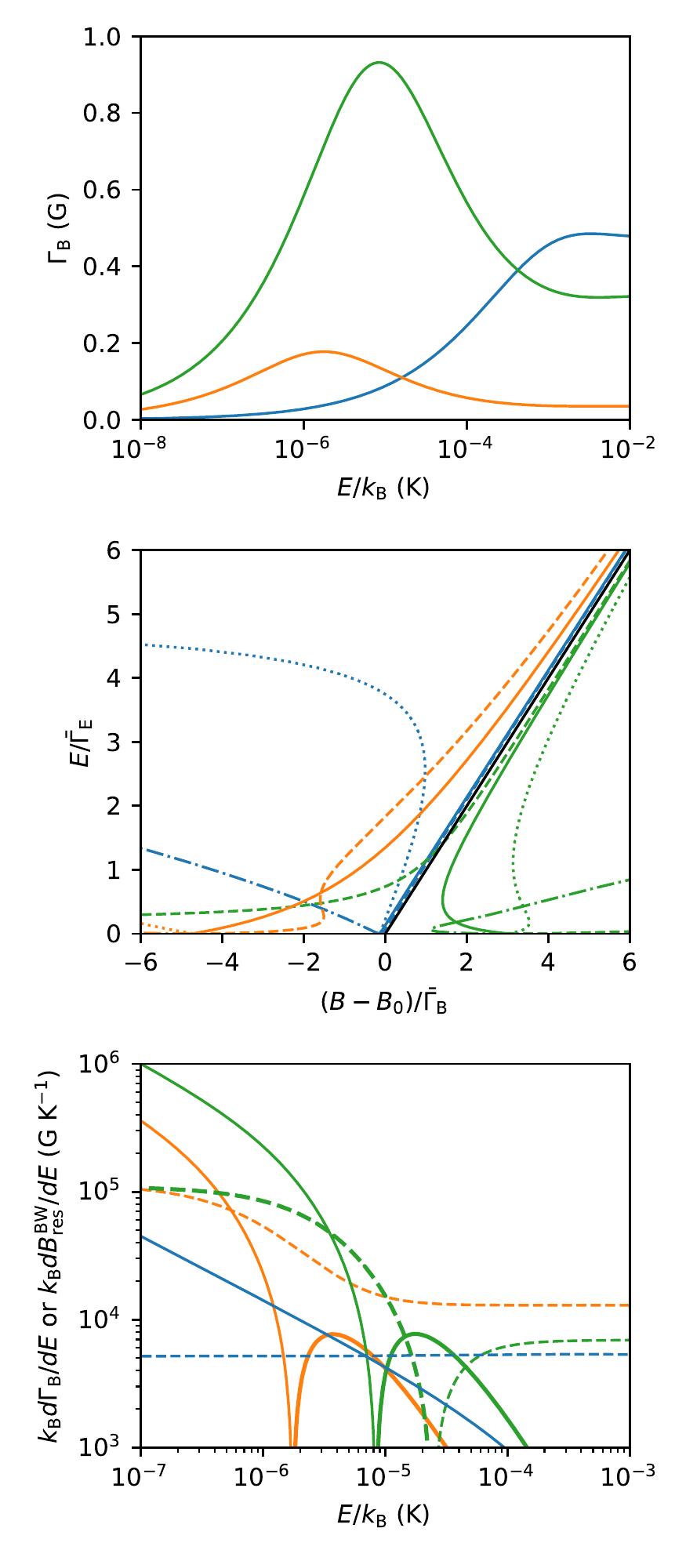}
\caption{\label{fig:params} Resonance parameters for resonances 1 (blue), 3
(orange), and 4 (green). (a) Width $\Gamma_\textrm{B}(E)$. (b)
$B_\textrm{res}^\textrm{BW}(E)$ (solid lines),
$B_\textrm{res}^\textrm{pole}(E)$ (dotted lines), and peak and trough in
$Q(E,B)$ (dashed and dot-dashed lines, respectively). The axes are scaled such
that the bare bound state (black line) coincides for all three resonances. (c)
Energy derivatives $d\Gamma_\textrm{B}/dE$ (solid lines) and
$dB_\textrm{res}/dE$ (dashed lines); thicker lines show the absolute values
when the quantities are negative.}
\end{figure}

Figure \ref{fig:params} shows the resonance parameters $\Gamma_\textrm{B}(E)$
and $B_\textrm{res}^\textrm{BW}(E)$ obtained from Eqs.\ \eqref{eq:QDT_Gamma} to
\eqref{eq:QDT_B0} for resonances 1, 3, and 4, together with their energy
derivatives. Panel (a) shows the width $\Gamma_\textrm{B}(E)$; the shapes are
the same as those of the corresponding functions $C^{-2}(E)$ in Fig.\
\ref{fig:qdt_params}. $\Gamma_\textrm{B}(E)$ for resonances 3 and 4 is greatly
enhanced around a few $\mu$K before reducing at higher energies. The
high-energy limits are the values of
$\bar{\Gamma}_\textrm{B}$; this is larger for resonance 1 than
for resonance 4, even though $\Delta$ is a factor of 6 smaller for resonance 1.
Similarly, the high-energy limit of $\Gamma_\textrm{B}(E)$ is much smaller for
resonance 3 than for resonance 1, even though $\Delta$ is similar for these two
resonances. These effects arise because $|a_\textrm{bg}|$ is large for
resonances 3 and 4, enhancing their widths near threshold.

The solid colored lines in Figure \ref{fig:params}(b) show
$B_\textrm{res}^\textrm{BW}(E)$ for the same three resonances; the axes are
scaled by the short-range widths $\bar{\Gamma}_\textrm{B}$ and
$\bar{\Gamma}_\textrm{E} = \mu_\textrm{rel} \bar{\Gamma}_\textrm{B}$, so that
the bare bound states coincide (black line). For resonance 1,
$B_\textrm{res}^\textrm{BW}(E)$ is close to the bare bound state; this is
because $a_\textrm{bg}$ is close to $\bar{a}$ so that $\tan \lambda (E)$ is
always small. For resonances 3 and 4, the deviations are much larger due to the
large magnitudes of $a_\textrm{bg}$ and resulting large $\tan\lambda (E)$. The
slopes $dB_\textrm{res}^\textrm{BW}/dE$ are constant in the low-energy limit,
but very different from that of the bare bound state. For resonance 4 the slope
at low energy has opposite sign to that at high energy because $a_\textrm{bg}$
is large and negative. The differences between $B_\textrm{res}^\textrm{BW}(E)$
and the position of the bare bound state descrease rapidly with energy

The dashed and dot-dashed lines in Figure \ref{fig:params}(b) show the
positions of the peaks and troughs in the time delay $Q(E,B)$, respectively
\footnote{Specifically, we solve $\partial Q/\partial B = 0$ for $B$ at fixed
$E$.}. These both coincide with $B_\textrm{res}^\textrm{BW}(E)$ at zero energy,
but they separate rapidly with increasing energy, by a quantity proportional to
$E^{3/2}$. For resonance 1 the peak is very close to
$B_\textrm{res}^\textrm{BW}(E)$; for resonance 3 it still approaches
$B_\textrm{res}^\textrm{BW}(E)$ at high energy, though in a more complicated
manner. For both these resonances the trough moves quickly away from
$B_\textrm{res}^\textrm{BW}(E)$ as it becomes shallow and unimportant. For
resonance 4 it is the trough that remains near $B_\textrm{res}^\textrm{BW}(E)$
at low energy, while the peak moves quickly away to high field and loses its
identity. A new peak then approaches from the low-field side and replaces the
trough near $B_\textrm{res}^\textrm{BW}(E)$.

The dotted lines in Figure \ref{fig:params}(b) show the positions
$B_\textrm{res}^\textrm{pole}(E)$ of the poles in the scattering length. They
coincide with $B_\textrm{res}^\textrm{BW}(E)$ at zero energy, but move away
from it rapidly as energy increases. They are unrelated to the peaks and
troughs in the time delay. They exhibit divergences in field as a
function of energy; these occur because $a_\textrm{bg}$ has a pole at every
energy where the background phase shift $\delta_\textrm{bg}(E)$ passes through
$(2n+1)\pi/2$, and $B_\textrm{res}^\textrm{pole}(E)$ undergoes a series of
avoided crossings with these background poles.

Figure \ref{fig:params}(c) shows the energy derivatives of
$B_\textrm{res}^\textrm{BW}(E)$ and $\Gamma_\textrm{B}(E)$. At low energy,
$dB_\textrm{res}^\textrm{BW}/dE$ is constant and $d\Gamma_\textrm{B}/dE$ is
proportional to $E^{-1/2}$. For resonance 1 they remain close to their high-
and low-energy limiting forms respectively, validating the approximations used
to derive Eq.\ \eqref{eq:crossover} for $E_\textrm{X}$; the two derivatives
cross near $7\ \mu\textrm{K}$ as predicted. For resonances 3 and 4, however,
$dB_\textrm{res}^\textrm{BW}/dE$ is dominated by threshold effects at low
energy. The functions do not approach their high-energy behavior
$dB_\textrm{res}^\textrm{BW}/dE=1/\mu_\textrm{rel}$ until $10\ \mu\textrm{K}$
or more. By contrast, $d\Gamma_\textrm{B}/dE$ deviates from its low-energy
limiting behavior well below $10\ \mu\textrm{K}$. Both approximations used to
derive Eq.\ \eqref{eq:crossover} thus break down for these resonances. The
behavior of $Q(E,B)$ can nevertheless be understood from the
derivatives in Figure \ref{fig:params}(c). In particular, it may be seen that
for resonance 3 there is a single crossover between
$dB_\textrm{res}^\textrm{BW}/dE$ and $d\Gamma_\textrm{B}/dE$, so that there is
a well-defined value of $E_\textrm{X}$, even though it is poorly approximated
by Eq.\ \eqref{eq:crossover} in this case. For resonance 4,
however, there are multiple crossovers and $E_\textrm{X}$ is poorly defined.

\section{Conclusions}

We have studied the behavior of the collisional time delay in cold and
ultracold atomic and molecular collisions. We have carried out coupled-channels
scattering calculations on ultracold collisions of $^{87}$Rb, $^{85}$Rb and
$^{133}$Cs in the vicinity of magnetically tunable Feshbach resonances. Far
above threshold, the time delay as a function of either energy or applied field
exhibits a symmetric peak whose integral is independent of the resonance width.
In the low-energy limit, however, the time delay is proportional to the
scattering length (and inversely proportional to the collision velocity or wave
vector $k$). Across a resonance, the scattering length passes through a pole as
a function of applied field; there are regions of large positive and large
negative time delay, and the resonant contribution averages to zero when
integrated over the field.

For resonances that are narrow, or have a background scattering length
$a_\textrm{bg}$ close to the mean scattering length $\bar{a}$, the transition
from the low-energy oscillation to the high-energy peak occurs around a
crossover energy $E_\textrm{X}$ that is proportional to the square of the
dimensionless resonance strength parameter $s_\textrm{res}$. For broad
resonances where $a_\textrm{bg}$ is large (either positive or negative), the
behavior is more complex.

The behavior of the time delay at low energy arises from the variation of the
resonance position and width near a scattering threshold. We have presented an
analysis based on multichannel quantum defect theory. For narrow resonances and
when $a_\textrm{bg} \approx \bar{a}$, the resonance position depends nearly
linearly on energy and the main threshold effect is from the energy dependence
of the resonance width. For broad resonances with $a_\textrm{bg} \gg \bar{a}$
or $a_\textrm{bg} \ll \bar{a}$, however, there are important additional effects
due to the effect of the threshold on the resonance position.

The results obtained here will be conceptually important in understanding
complex formation during ultracold collisions, which is believed to play an
important role in losses of nonreactive molecules from traps. The resonances
considered have comparable short-range widths to those predicted for collisions
of ultracold molecules \cite{Christianen:density:2019}.

\begin{acknowledgments}
We are grateful to Ruth Le Sueur for valuable discussions. This work was
supported by the U.K. Engineering and Physical Sciences Research Council
(EPSRC) Grants No.\ EP/N007085/1, EP/P008275/1 and EP/P01058X/1.
\end{acknowledgments}

\bibliography{../../all,timedelays}

\end{document}